\documentclass{egpubl}
\usepackage{eurova2026}

\WsPaper           % uncomment for final version of EG Workshop contribution
\usepackage[T1]{fontenc}
\usepackage{dfadobe}  

\usepackage{cite}  % comment out for biblatex with backend=biber
\BibtexOrBiblatex
\electronicVersion
\PrintedOrElectronic
\ifpdf \usepackage[pdftex]{graphicx} \pdfcompresslevel=9
\else \usepackage[dvips]{graphicx} \fi

\usepackage{egweblnk} 
\newcommand{\paperTitle}{Visual Boosting Techniques for Spatiotemporal Dense Pixel Visualizations}
\usepackage{svg}
\usepackage{xspace}
\usepackage[subtle]{savetrees}
\usepackage[most]{tcolorbox}

\usepackage[english]{babel}
\addto\extrasenglish{

}

\newcommand{\subhead}[1]{\noindent \textbf{#1 --}}

\newcommand\figVSpace{-1.5em}

\newtcbox{\iconbox}{
  on line,
  box align=base,
  colback=white,
  colframe=gray,
  arc=.1ex,
  boxsep=0.3ex,
  left=0pt,
  right=0pt,
  top=0pt,
  bottom=0pt
}

\newcommand{\boosticonboxed}[1]{%
  \iconbox{\includesvg[height=.75em]{#1}}\xspace%
}

\newcommand{\Bhatching}{\boosticonboxed{figs/inline/hatching}}
\newcommand{\Bdistortion}{\boosticonboxed{figs/inline/distortion}}
\newcommand{\BmapGlyph}{\boosticonboxed{figs/inline/map}}
\newcommand{\Bhalo}{\boosticonboxed{figs/inline/halo}}
\newcommand{\Bcolor}{\boosticonboxed{figs/inline/color}}

\title[\paperTitle]%
      {\paperTitle}

\author[J. Rauscher et al.]
{\parbox{\textwidth}{\centering 
        Julius Rauscher$^1$\orcid{0000-0003-1318-9642},
        Frederik L. Dennig$^1$\orcid{0000-0003-1116-8450},
        Udo Schlegel$^2$\orcid{0000-0002-8266-0162},
        Daniel A. Keim$^1$\orcid{0000-0001-7966-9740},
        and Tobias Schreck$^3$\orcid{0000-0003-0778-8665}
        }
        \\
{\parbox{\textwidth}{\centering 
$^1$University of Konstanz, Germany;
$^2$LMU \& MCML Munich, Germany;
$^3$TU Graz, Austria
}
}
}

\begin{document}

% uncomment for using teaser
\teaser{
 \includegraphics[width=\linewidth]{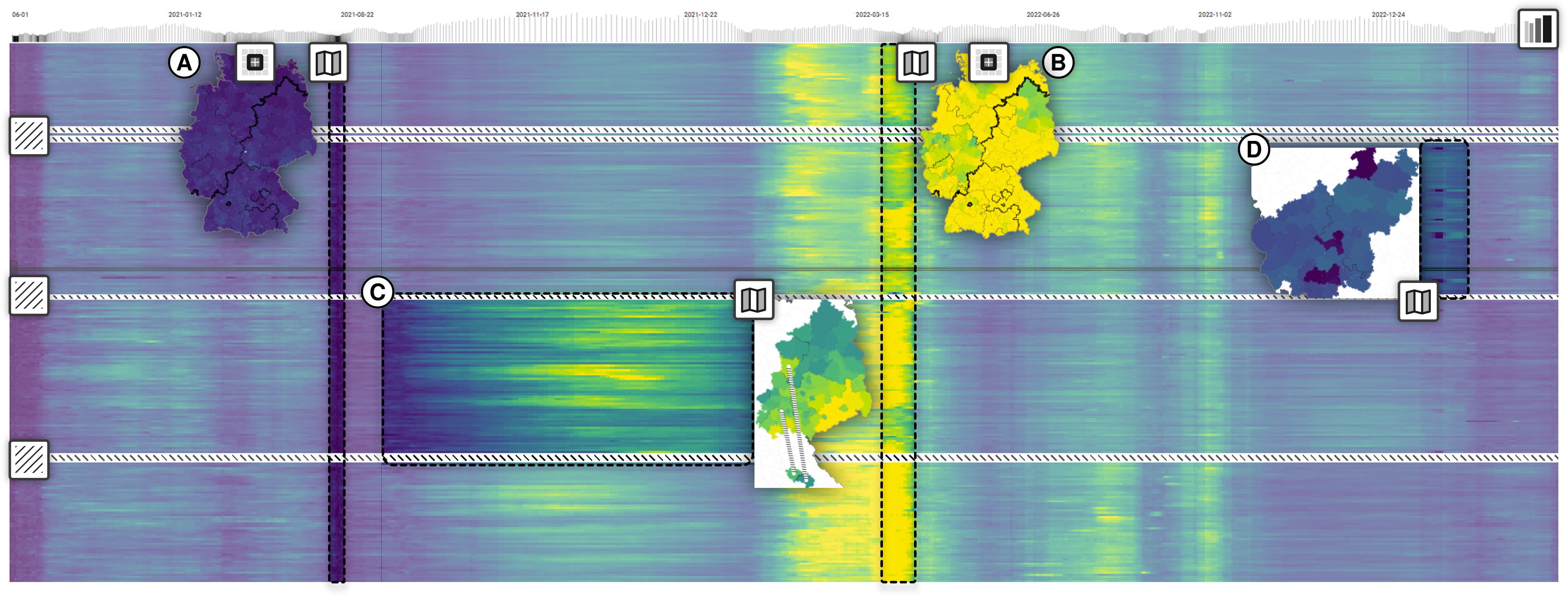}
 \centering
  \caption{
  Overview of COVID-19 incidence data of German districts from 2020 to 2023 using a dense pixel visualization, where rows resemble geographic locations and columns represent time.
  The column widths are scaled by spatial autocorrelation, emphasizing points in time when spatial clusters form.
  To expose linearization artifacts, large geographical distances in the ordering are indicated by \emph{\Bhatching Trustworthiness Gaps}. 
  By brushing the dense pixel view, \emph{\BmapGlyph Map Glyphs} can be rendered to assess the geographical coherence, where \emph{\Bhalo Discontinuity Borders} indicate large ordering distances of geographical neighbors.
  }
\label{fig:teaser}
}

\maketitle
%-------------------------------------------------------------------------
\begin{abstract}
The analysis of spatiotemporal data is essential in domains such as epidemiology and environmental monitoring, where understanding the interplay between spatially distributed phenomena and their temporal evolution is critical. 
Dense pixel visualizations offer a compact, effective overview of spatiotemporal dynamics.
However, the necessary linearization of 2D geographic space into a 1D ordering inevitably introduces structural distortions that manifest as visual artifacts.
We propose a measure-driven visual analytics approach that captures visual artifacts through neighborhood preservation measures for 1D orderings and renders them using visual boosting techniques such as glyphs, halos, and hatching.
We demonstrate our approach through a usage scenario analyzing COVID-19 incidence data across German districts, showing that interactive, measure-driven boosting enables analysts to reliably distinguish genuine spatial patterns from linearization artifacts.

%The tool at \url{http://dl.acm.org/ccs.cfm} can be used to generate
% CCS codes.
%Example:


\end{abstract}

\section{Introduction}
Geolocated time series are central in domains such as epidemiology~\cite{slingsby:gridmapcovid:2023} and environmental monitoring~\cite{diehl:vaWeatherForecast:2015,deng:geochron:2024}, but are difficult to analyze because they require simultaneous reasoning about evolving spatial configurations and local temporal variation~\cite{spaceTimeTimeSpace:andrienko:10}.
Dense pixel visualizations provide compact overviews for collective movement~\cite{motionrugs:buchmueller:19, spatialrugs:buchmüller:2021} or spatio-temporal events such as wildfires~\cite{1dva:franke:21} or storms~\cite{koepp:tempMergeTreeMap:2023}, 
where values are encoded as colored pixels arranged in a matrix-like layout with one axis representing space and the other time.
In contrast to animation, such representations are not susceptible to change blindness and, unlike 3D-based approaches (e.g., Space-Time Cubes~\cite{stcVolumes:deng:2025}), they avoid issues of occlusion and perspective distortion, thereby supporting consistent visual comparison across space and time.
However, the expressiveness and effectiveness of such a dense pixel arrangement depend heavily on the ordering of spatial entities. 
The linearization from 2D geographical space to a 1D ordering inevitably introduces distortions, generating visual artifacts in the visualization~\cite{motionrugs:buchmueller:19, koepp:tempMergeTreeMap:2023} that can prompt false conclusions about the underlying data.

Such distortions can be quantified through neighborhood preservation in both the spatial and ordering domain, and various quality measures have been proposed for geospatial point~\cite{ksss:guo:06} or polygon~\cite{rauscher:1d-order-poly:2025} data.
These quality measures have been used to either assess the quality~\cite{stablevissum:wulms:21} or to determine suitable ordering strategies~\cite{1dva:franke:21}.
Yet for complex geographies, distortions cannot be entirely eliminated, regardless of the chosen strategy. 
This limitation highlights not only the need to measure ordering quality but also the need to visually communicate shortcomings.
Visual boosting techniques~\cite{visualBoosting:oelke:11} offer a promising direction, as they can selectively emphasize regions of interest in dense pixel displays without altering the underlying data encoding.

In this paper, we propose a measure-driven approach that integrates established ordering quality metrics with visual boosting techniques, such as glyphs, halos, hatching, and distortion, to enhance dense pixel visualizations for geolocated time series data.
Building on prior work in visual boosting, we investigate how metric-based information about neighborhood distortions can be embedded directly into the visualization to make structural limitations salient.
Thereby, we make the following contributions:
\begin{itemize}%
    \item A \href{https://1d-poly-order.dbvis.de/}{\textbf{Visual Analytics prototype}} for spatiotemporal data, integrating visual boosting techniques based on quality measures to make linearization artifacts salient.
    \item A \textbf{Usage Scenario} on COVID-19 data in Germany, showcasing how measure-driven boosting supports reliable pattern assessment.
\end{itemize}

\section{Related Work}

\subhead{Spatiotemporal Data Visualization}
Developments across space and time are commonly visualized using multiple linked views, combining a map with timelines~\cite{diehl:vaWeatherForecast:2015}, calendar views~\cite{Li:vismate:2014}, or storylines~\cite{deng:geochron:2024}.
While these approaches benefit from dedicated representations for each dimension, space and time are juxtaposed in distinct views, creating a visual discontinuity.
Glyphmaps integrate the spatial and temporal components into a single view, either in a gridded~\cite{wickham:gylphmaps:2012, slingsby:gridmapcovid:2023} or entity-centered layout~\cite{mcnabb:MultivariateMapsGlyphPlacement:2019}, yet their scalability is limited by the grid layout or glyph size.
3D-based visualization approaches, such as 3D timelines on maps~\cite{3DtimeSeriesMap:Thakur:2010}, stacked trajectory bands~\cite{tominski:stackedTrajectories:2012}, or Space-Time cubes~\cite{stcVolumes:deng:2025}, integrate the temporal information as a third axis, but suffer from perspective distortion and occlusion.
Despite the variety of the above-mentioned approaches, they all struggle with providing a compact overview of spatiotemporal developments at scale.
Dense-pixel visualizations offer a promising solution for overview tasks, as they compactly map individual data values to single pixels~\cite{keim:pixelOriented:2000}, enabling the display of large spatial and temporal extents within a single coherent view.

\subhead{Dense Pixel Visualizations}
To investigate the dynamics of collective animal behavior, Buchmüller et al. use a Hilbert space-filling curve to linearize space and construct dense pixel layouts with space along the y-axis and time on the x-axis, where color can be used to either encode data attributes~\cite{motionrugs:buchmueller:19} or spatial locations~\cite{spatialrugs:buchmüller:2021}.
As subsequent time instances can produce vastly different orderings, Wulms et al.~\cite{stablevissum:wulms:21} propose using Stable Principal Components to improve ordering stability and mitigate visual artifacts.
Stolk et al.~\cite{groupRugs:stolk:2025} leverage additional whitespace between moving groups to visually indicate clusters.
Aside from moving entities, Franke et al.~\cite{1dva:franke:21} show how
spreading events such as wildfires or pandemics can be visualized using dense pixel visualizations.
Köpp and Weinkauf~\cite{koepp:tempMergeTreeMap:2023} apply it to scalar fields by using a feature-based linearization based on augmented merge trees.
Zhou et al.~\cite{datadrivensfc:zhou:21} propose data-driven space-filling curves for multiscale data visualization in regular grids by balancing feature and locality coherence.
%Position plots, position of soccer players over time~\cite{brandes:shapegraphs:2025}.
Further applications of dense pixel views in other domains include dynamic graphs~\cite{cui:letflow:2014, dg2pix:cakmak:2020}, sensor networks~\cite{pham:contimap:2020}, machine learning model attributions~\cite{schlegel:densePix:2023}, moving regions~\cite{morevis:valdrigighi:24}, or physical activity data~\cite{motivator:rauscher:2025}.

While dense pixel visualizations provide a compact overview of large amounts of spatiotemporal data, the linearization of 2D space inevitably produces visual artifacts, labeled as phantom splits~\cite{motionrugs:buchmueller:19}, inconsistent visual patterns~\cite{1dva:franke:21}, or temporal discontinuities~\cite{koepp:tempMergeTreeMap:2023}, necessitating designated quality measures to locate and quantify the impact of these artifacts. 
Oelke et al.~\cite{visualBoosting:oelke:11} demonstrate that visual boosting techniques such as halos, coloring, distortion, and hatching can be used to accentuate interesting or important datapoints in pixel-based visualizations.
Conversely, we argue that such techniques are just as valuable to 
communicate structural limitations and make visual artifacts more salient through the embedding of quality measures.

\subhead{Quality Measures}
The assessment of neighborhood preservation through quality metrics has been extensively studied in the context of dimensionality reduction~\cite{jeon:drsurvey:2025}.
For such projections, Venna and Kaski~\cite{m1m2:venna:01} define \emph{Trustworthiness} and \emph{Discontinuity} measures to quantify whether ordering neighbors are spatially close and vice versa.
Guo and Gahegan~\cite{ksss:guo:06} provide analogous distance- and rank-weighted variants tailored towards point-based spatial data.
Rauscher et al.~\cite{rauscher:1d-order-poly:2025} extend these measures to contiguous polygon data by introducing locally varying neighborhood sizes and contiguity-based distance functions.
While such measures have been employed to compare and determine suitable ordering strategies~\cite{motionrugs:buchmueller:19,stablevissum:wulms:21, 1dva:franke:21}, their local error scores capture the presence of artifacts and can be further used to visually highlight these shortcomings.

\begin{figure*}[t!]
    \centering
    \includesvg[inkscapelatex=false, inkscapedpi=600,width=\linewidth]{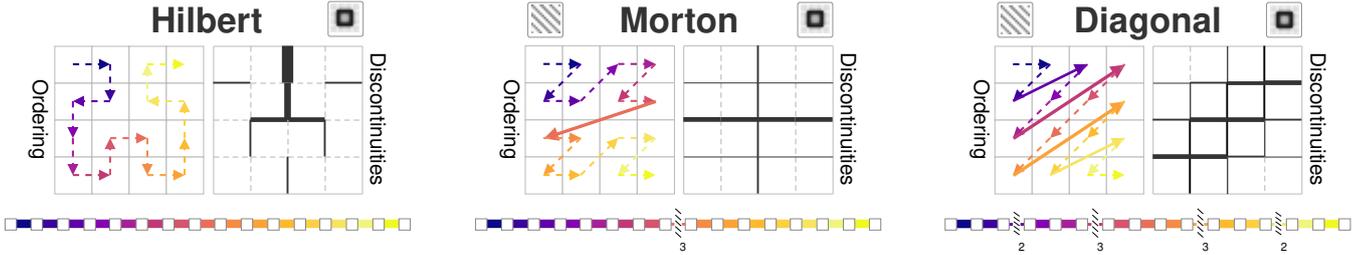}
    \caption{Showing \emph{\Bhatching Trustworthiness Gaps} and  \emph{\Bhalo Discontinuity Borders} for different space-filling curves on a 4x4 grid. A Hilbert curve always follows geographic neighbors and creates no gap, but exhibits the largest border where geographic neighbors are distant in the ordering. A Morton curve yields one gap (assuming Queen contiguity) for $\beta<=3$, and its recursive grid is reflected in the borders. The Diagonal ordering introduces gaps for every diagonal jump, and exhibits larger borders in the center of the grid. 
    }
    \label{fig:trustgaps} 
    \vspace{\figVSpace}
\end{figure*}
\vspace{-1em}
\section{Visual Analytics Prototype}
Spatiotemporal data exists in various topological forms.
In this work, we focus on data defined by contiguous polygons, such as administrative regions, which serve as the primary unit for socioeconomic and epidemiological reporting.
Each polygon is spatially static and contains a value for every timestep, such as daily measurements.
Unlike point-based data, polygons allow semantically meaningful neighborhood definitions through contiguity. 
In contrast to moving entities, the fixed topology enables a single ordering and associated quality measures to be applied consistently across time.

To mitigate visual artifacts and make structural limitations more salient, we enrich our visualization with several visual boosting techniques inspired by Oelke et al.~\cite{visualBoosting:oelke:11}.

\begin{figure}[t!]
    \centering
    \includegraphics[width=\linewidth]{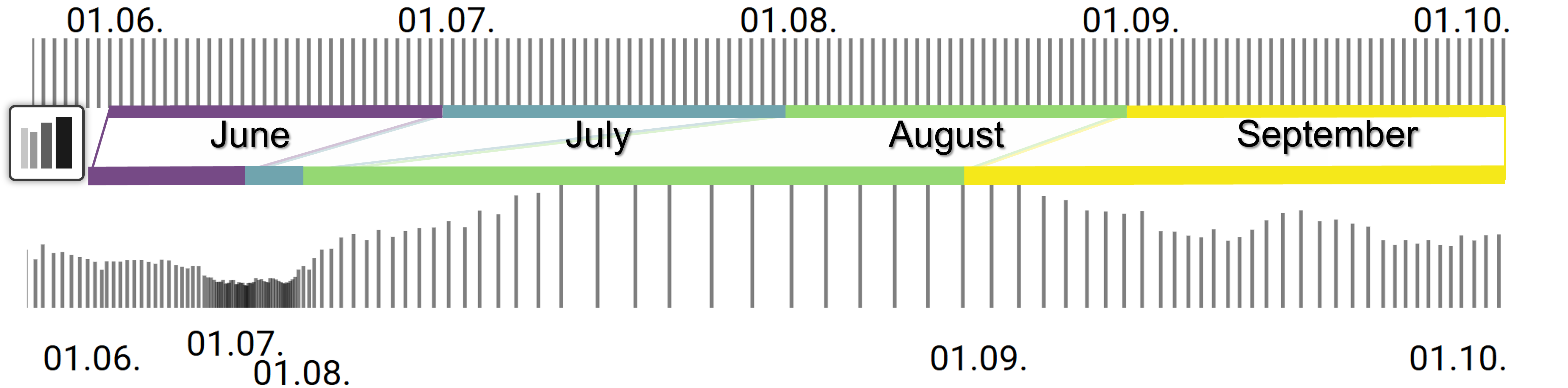}
    \caption{Effect of \emph{\Bdistortion Boosting with Distortion} on the timeline and pixel width. Time instances with lower spatial autocorrelation are shrunk (June--July), and those with higher values are enlarged (August--September), further indicated by the tick height. }
    \label{fig:timeline} 
    \vspace{\figVSpace}
\end{figure}

\begin{figure*}
    \centering
    \includegraphics[width=\linewidth]{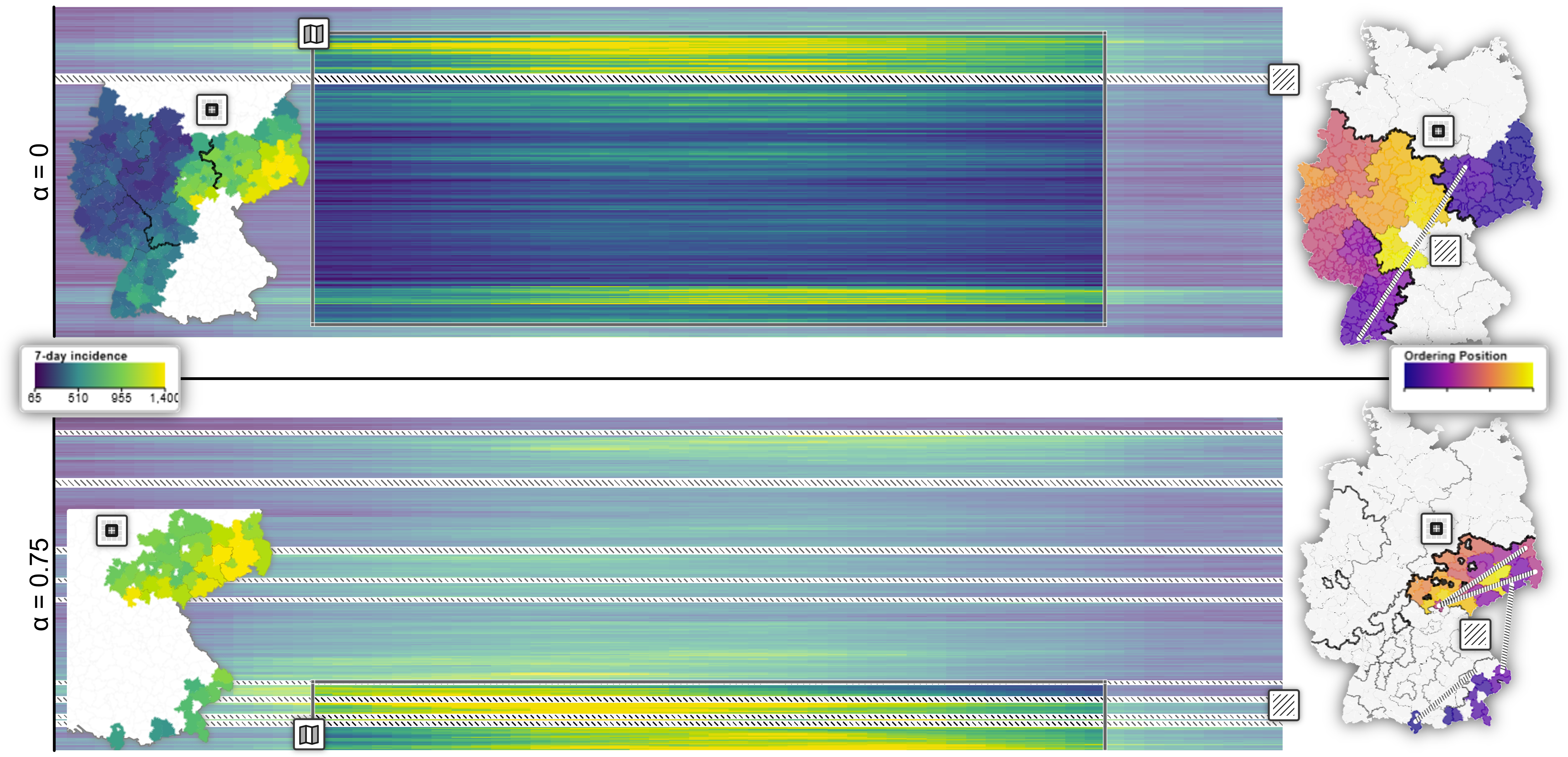}
    \caption{Using different $\alpha$ to determine orderings (with $\beta = 5$). For $\alpha=0$ (top), we only observe one \emph{\Bhatching Trustworthiness Gap}, however, the \emph{\Bhalo Discontinuity Borders} in the \emph{\BmapGlyph Map Glyph} confirm that a spatial cluster of high values is split into two locations in the dense pixel layout. Using $\alpha = 0.75$ (bottom), that geographic region is more compactly represented in the ordering, at the cost of additional \emph{Trustworthiness Gaps}.}
    \label{fig:alpha}
        \vspace{\figVSpace}
\end{figure*}

\subsection{Dense Pixel Visualization}

\subhead{Ordering}
In our dense pixel visualization, rows represent spatial entities and columns time steps, with each pixel encoding the value of a geographic entity at a given timestep. Pixel colors are mapped using the viridis colorscale, implicitly introducing \emph{\Bcolor Boosting with Color}.
While traditional approaches aimed to preserve geographical coherence in the row ordering~\cite{motionrugs:buchmueller:19, spatialrugs:buchmüller:2021, stablevissum:wulms:21}, more recent work has highlighted the benefits of also integrating temporal similarity~\cite{1dva:franke:21, koepp:tempMergeTreeMap:2023}.
As both variants have benefits and drawbacks depending on the analysis task, we introduce a parameter $\alpha$ that weights standardized pairwise distances from the geographic domain using centroid distances ($d_{geo}$) and from the timeseries using Euclidean distance ($d_{ts}$):
\begin{equation}
    D_{mix}(x_i, x_j) = (1-\alpha) \cdot \frac{d_{geo}(x_i,x_j) - \overline{d}_{geo}}{\sigma(d_{geo})} + \alpha \cdot \frac{d_{ts}(x_i,x_j) - \overline{d}_{ts}}{\sigma(d_{ts})}
    \label{eq:distance-fn}
\end{equation}
Using this distance function with agglomerative hierarchical clustering (AHC), we obtain an \emph{optimal} 1D ordering from the leaf nodes of the dendrogram, as proposed by Guo and Gahegan~\cite{ksss:guo:06}.
Rauscher et al.~\cite{rauscher:1d-order-poly:2025} indicate that linkage criteria  (except single linkage) produce qualitatively similar orderings. We select Ward~\cite{wardlinkage:ward:1963}, whose minimum-variance criterion favors compact, homogeneous clusters.

\subhead{\Bdistortion Temporal Distortion}
Due to their space-efficient design, dense pixel views provide high information density, making them especially suitable for overview tasks. 
However, when analyzing developments over long time periods, individual extreme values can dominate the color scale and compress the dynamic range of the remaining data, thereby obscuring subtle temporal trends.
Similar to Hao et al.~\cite{hao:tsDistortion:2007}, we apply \emph{Boosting with Distortion} to emphasize time periods of interest.

Spatial autocorrelation directly reflects how strongly similar values concentrate among neighboring polygons at a given timestep, which is why we compute the standardized global Moran's~I~\cite{moran:i:1950} at each timestep, and use this value to scale the column widths of our dense pixel view.
This allocates greater screen space to time steps that deviate most strongly from the temporal baseline.
Oelke et al. noted that distortions can \emph{``decrease the user's ability to follow the course of values''}~\cite[p.879]{visualBoosting:oelke:11}, which is especially relevant for our temporal distortion that disrupts the linear flow of time.
To increase the visibility of the distortion, we further scale the height of the timeline ticks, yielding a barchart-like profile whose height and spacing mirror the distorted column widths (see \autoref{fig:timeline}).

\subhead{\Bhatching Trustworthiness Gaps}
Linearizing 2D space is an error-prone, lossy dimensionality-reduction process. 
A major shortcoming is that geographically distant entities may become ordering neighbors, which has been previously quantified as a \emph{Trustworthiness Error}~\cite{m1m2:venna:01,rauscher:1d-order-poly:2025}.
Such geographic jumps in the ordering sequence can be further introduced when integrating timeseries similarity into the distance function (see \autoref{eq:distance-fn}). 
We indicate these errors by drawing horizontal gaps in the dense pixel view, effectively dissecting the visualization into geographically coherent clusters (similar to GroupRugs~\cite{groupRugs:stolk:2025}).
Inspired by \emph{Boosting with Hatching} technique~\cite{visualBoosting:oelke:11}, we use a hatching pattern to visually distinguish them from rows displaying actual data values. 

To determine a Trustworthiness error, we construct an undirected adjacency graph $G = (V,E)$, where each vertex $ v \in V$ represents a polygon and an edge $(u,v) \in E$ indicates geographical contiguity through a shared border.
Let $(v_1,...,v_N)$ denote the imposed ordering of the vertices.
For each consecutive pair $v_n, v_{n+1}$ in the ordering, we compute their shortest-path distance in $\delta_{G}(v_n, v_{n+1})$, corresponding to the minimal number of contiguous regions that must be traversed to connect the two polygons.
If this distance exceeds a given integer threshold \boldmath$\beta$, we mark this as a Trustworthiness violation and indicate this in the dense pixel view with a gap.
\begin{equation}
    \epsilon_n = \mathbf{1} [ \delta_G(v_n, v_{n+1}) > \beta ],\qquad n = 1, ..., N-1
\end{equation}
In contrast to previous measures~\cite{ksss:guo:06, rauscher:1d-order-poly:2025}, this integer conveys semantic meaning and is hence more suitable for an interactively steerable parameter. \autoref{fig:trustgaps} exemplifies the determination. 

\subsection{Geographic Map}
\subhead{\BmapGlyph Map Glyph}
Additional geographic maps can be overlaid on the dense-pixel view to summarize spatiotemporal extent, inspired by \emph{Boosting with Glyphs}.
Brushing over the dense pixel view creates a selection for which a \emph{Map Glyph} is shown adjacent to the brushed region (see \autoref{fig:alpha}).
The map keeps the initial geographic extent, but only renders polygons included in the selection.
The corresponding data values are visualized through interactively changeable statistical aggregations such as min, mean, and max.

\subhead{\Bhalo Discontinuity Borders}
Aside from Trustworthiness Errors where ordering neighbors are spatially distant, 1D orderings can furthermore exhibit Discontinuities where geographical neighbors are placed at distant indices in the ordering.
This is not directly apparent in the dense pixel layout, but may lead to a scattering of a spatially coherent pattern (see \autoref{fig:alpha}). 
Akin to~\cite{rauscher:1d-order-poly:2025}, we can determine the ordering distance for every polygon pair that shares a contiguous border. 
Oelke et al.~\cite{visualBoosting:oelke:11} hinted at the potential of \emph{Boosting with Halos} in geospatial analysis. 
To highlight spatial ordering discontinuities, we use the polygon boundaries in the geographic map as halos and scale their stroke widths by the ordering distance, making larger distances more visually dominant.
This provides a visual indication of ordering quality, with more borders indicating a more geographically fragmented ordering. \autoref{fig:trustgaps} exemplifies this for different space-filling curves.

\subhead{Ordering Path}
To enhance the spatial understanding of the ordering, its geographic traversal can be shown on the \emph{Map Glyph} on demand (see \autoref{fig:alpha}).
Each polygon is colored according to its position in the ordering using a sequential color scheme, indicating discontinuities by color differences.
A link corresponding to a \emph{\Bhatching Trustworthiness Gap} is drawn as a hatched line to maintain visual consistency with the dense pixel view.

\section{Usage Scenario} %Evaluation
To demonstrate the capabilities of the prototype, we present a usage scenario on daily COVID-19 incidence data recorded in 400 German districts from 2020 to 2024.

\subsection{Overview}

\autoref{fig:teaser} shows the full extent of the data, exhibiting a general ripple-like pattern of multiple waves of globally low or high values, reflecting seasonal fluctuations common for respiratory illnesses (\autoref{fig:teaser} A\&B). 
Furthermore, several local outliers in some districts can be identified where cases drop and then rise again, likely due to a delay in number reporting.
By brushing over the region in the dense pixel view, a \emph{Map Glyph} reveals the spatial extent of these outliers, and confirms that they are spatially close, although not all contiguous (\autoref{fig:teaser} D).

To identify local spatial clusterings, the user applies \emph{Boosting by Distortion} to rearrange the temporal domain and accentuate time instances where the Moran's~I measure indicates strong spatial autocorrelation.
This causes time instances with random or uniform spatial distribution (such as the global waves) to shrink (\autoref{fig:teaser} A\&B), whereas time intervals preceding the waves are enlarged, especially accentuating the period between November 2021 to January 2022 (\autoref{fig:teaser} C).
To better understand the dynamics of this interval, the user updates the temporal reference extent, triggering a recomputation of the ordering based on local timeseries similarity, which results in a reevaluation of the \emph{Trustworthiness Gaps} and \emph{Discontinuity Borders}.

\subsection{Measure-driven Exploration}

\autoref{fig:alpha} shows the selected timeframe from 01.11.2021 to 01.01.2022 using two orderings obtained from different $\alpha$ values with $\beta = 5$.
When ordering exclusively with respect to geographical similarity ($\alpha = 0$), only a single \emph{Trustworthiness Gap} is drawn, indicating that the ordering avoids large geographical jumps.
However, the \emph{Discontinuity Borders} on the map indicate two larger rifts where geographical neighbors are distant in the ordering.
This results in the dispersion of a spatially coherent cluster in the Eastern regions of Germany, which can be verified by brushing over the two patterns. 
This observation aligns with prior reports of an East–West separation in incidence dynamics.~\cite{covidgermany:diebner:2023}.

When adjusting the ordering to incorporate more temporal similarity ($\alpha = 0.75$), we can observe that the previously divided pattern has merged and is contained in a smaller region cluster located at the bottom of the view.
When inspecting the pattern using the \emph{Map Glyph}, the geographical coherence is further reflected by the \emph{Discontinuity Borders}, which now better shape the geographical extent of the spatiotemporal cluster.
However, multiple \emph{Trustworthiness Gaps} are introduced, indicating that the ordering exhibits geographical jumps within the subextent.
Additionally, several \emph{holes} are observable in the geographical extent, which do not seem to follow the surrounding trend.
Upon closer examination, it becomes apparent that these holes correspond to the larger cities in the region, which exhibit a different temporal development compared to the more rural districts, and are therefore more distant in the ordering.

\section{Discussion}
\subhead{Ordering}
The flexible distance function allows exploration of the balance between geographic and temporal perspectives.
Purely geographic orderings preserve spatial continuity but can split temporally coherent patterns, leading to larger \emph{Discontinuity Borders}.
Emphasizing timeseries similarity produces more temporally coherent pixel arrangements, at the cost of fragmenting the geography with \emph{Trustworthiness Gaps}.

\subhead{\Bcolor Boosting with Color}
Our design incorporates a global colorscale in the dense pixel visualization that is also used to color the \emph{Map Glyphs}.
While this preserves the semantic meaning of color, smaller data ranges can result in less obvious visual patterns. 
For visual highlighting, we increase the lightness of pixels outside the brushed selected region.
Alternatively, a focus blur could be used to create a selection highlight, a technique commonly used to achieve a lens metaphor.

\subhead{\Bdistortion Boosting with Distortion}
Scaling the temporal axis by Moran's~I effectively guides the user towards time steps with spatially coherent patterns (see \autoref{fig:teaser} C). 
However, temporal instances with spatially noisy or heterogeneous distributions are compressed, which might be undesirable depending on the analysis task.
Other interestingness measures (e.g., standard deviation or entropy) could be explored, and alternative boosting techniques that retain the linear flow of time, such as color saturation or focus blurring, might also be viable options.

\subhead{\Bhatching Boosting with Hatching}
By visually separating geographically distant ordering neighbors, \emph{Trustworthiness Gaps} prevent the viewer from possibly drawing false conclusions about spatial proximity in the dense pixel view.
At the same time, they are sensitive to the threshold parameter~$\beta$, and a too-low threshold will fragment the visualization into many small clusters, rendering it unreadable.
We provide an intuitive interpretation of $\beta$ as the number of contiguous geographic hops, and a default value can be obtained heuristically, i.e., using the average centrality. 
The computation method is further sensitive to polygon size, as larger polygons lead to an underestimate of the true spatial distance. 
Beyond trustworthiness, hatching could be extended to depict multiple measures, each distinguishable through varying hatching styles or directions.

\subhead{\BmapGlyph Boosting with Glyphs}
The map glyphs provide a small spatiotemporal summary and support the spatial understanding of subregions in the dense pixel visualization at the cost of occluding parts of the dense pixel view.
Multiple glyphs can facilitate the comparison of different trends, where sophisticated placement algorithms are beneficial for an optimal alignment.
The proposed map glyph is limited to a choropleth map showing different statistical aggregation measures, and more complex designs could incorporate additional information.

\subhead{\Bhalo Boosting with Halos}
By scaling polygon border widths proportionally to ordering distance, the map view provides a spatial reference for interpreting fragmentation in the pixel layout.
As shown in \autoref{fig:alpha}, this can be beneficial to identify contiguous spatiotemporal patterns that have been scattered in the dense pixel visualization.

\subsection{Limitations and Future Work}
\subhead{Guidance}
The parameter configuration requires manual adjustment. While $\beta$ is intuitively defined as the number of contiguous geographic hops and can be heuristically initialized (e.g., by average centrality), optimal $\alpha$ and $\beta$ values depend on the data and task. 
Moreover, pattern detection relies on visual inspection and could be complemented by automated clustering or anomaly detection methods.

\subhead{Performance and Stability}
Modifying $\alpha$ requires recomputing the distance function, deriving a new ordering via AHC, and subsequently recalculating all quality measures that drive the visual boostings, which does not efficiently scale as the data size increases. 
Furthermore, AHC orderings are sensitive to small perturbations in the input, and a minor change in $\alpha$ can yield a substantially different ordering.

\subhead{Evaluation}
The usage scenario demonstrates the effectiveness of the proposed boosting methods, though examples from other domains would strengthen the tool's generalizability.
Furthermore, it remains unclear how target users perceive and interpret the boostings.
A controlled user study is needed to assess whether the techniques reliably support the intended analytical tasks and actually reduce misinterpretation of ordering artifacts. 
Such a study could systematically compare the effectiveness of boosting techniques, such as contrasting distortion against color, or evaluate the impact of different quality measures.

\section{Conclusion}

We presented a measure-driven visual analytics approach for dense pixel visualizations of geolocated time series data, bridging the inherent duality of spatiotemporal data. 
By embedding neighborhood preservation measures directly into the visualization using visual boosting techniques, we address the challenge that linearizing 2D geographic space inevitably introduces structural artifacts.
Our interactive approach makes ordering limitations explicitly visible and supports reliable and scalable spatiotemporal pattern exploration.
A usage scenario on COVID-19 incidence data across German districts demonstrates that measure-driven boosting effectively supports the visual assessment of spatiotemporal patterns in the presence of linearization artifacts.

\section*{Acknowledgement}
This work was funded by the Federal Ministry for Economic Affairs and Climate Action -- 03EI1048D.
\clearpage
%-------------------------------------------------------------------------

%-------------------------------------------------------------------------
% bibtex
\bibliographystyle{eg-alpha-doi} 
\bibliography{references/refs}       

% biblatex with biber
% \printbibliography                

%-------------------------------------------------------------------------
%Color tables are no longer required for purely electronic publications.
\newpage

\end{document}